Tuning the pH Response of Monolayer Hexagonal Boron Nitride/Graphene Field-Effect Transistors

*Nicholas E. Fuhr [a],\*, Mohamed Azize [a], David J. Bishop [a,b,c,d,e]*

*[a] Division of Materials Science and Engineering, Boston University, Boston, Massachusetts 02115, United States*

*[b] Department of Mechanical Engineering, Boston University, Boston, Massachusetts 02115, United States*

*[c] Department of Electrical and Computer Engineering, Boston University, Boston, Massachusetts 02115, United States*

*[d] Department of Biomedical Engineering, Boston University, Boston, Massachusetts 02115, United States*

*[e] Department of Physics, Boston University, Boston, Massachusetts 02115, United States*

[*] Correspondence: <u>fuhrnick@bu.edu</u>

## Abstract:

The chemical activity of ionized hydrogen describes the potency of protons in solution and is used to define the pH with wide-ranging applications in biological and materials sciences. Measuring pH with graphene field-effect transistors (FETs) has been described in the past but has yet to be characterized with a monolayer hexagonal boron nitride (hBN) capping layer. hBN capping is hypothesized to lower the chemical affinity of protons to oxidized defects in the graphene crystal and lower the standard deviation of pH measurement via screening charge density. First, the electronic properties of commercial, monolayer graphene and monolayer hBN/graphene, both on four-inch 90 nm $SiO_2$/p-type Si, were contrasted as a function of solutal pH in 10 mM phosphate buffered solution. The two-dimensional (2D) FETs were fabricated with photoresistless metallization followed by microcentrifuge tube-masking and reactive ion etching to define a quasi-pure 2D sensing mesa for pH sensing. Thereafter, the devices were exposed to phosphate buffer of varying pH and the resistance and liquid-transfer characteristics measured. The Dirac voltages of both devices became less positive with increasing pH. The hBN/graphene/$SiO_2$ FETs had higher linearity, and both lower standard deviation and range of Dirac voltage shifting than monolayer graphene/$SiO_2$. Then, before microcentrifuge tube-masking, freshly metallized devices were coated with 9 nm of $Al_2O_3$ by electron beam deposition. The $Al_2O_3$/graphene/$SiO_2$ and $Al_2O_3$/hBN/graphene/$SiO_2$ devices showed an altered response to the pH of the phosphate buffered solutions before and after basic etching of the $Al_2O_3$ with NaOH (pH = 12). This prompted studying the pH-dependencies of the 2D devices as a function of the thickness of $Al_2O_3$ via atomic layer deposition. These results show that modifying the surface of monolayer graphene with nanoscale dielectrics (monolayer hBN and/or $Al_2O_3$) enables tuning of the pH-dependent electronic properties of graphene pH sensors.

## Introduction

Protons are positively-charged nuclides that are present in solutions as ionized hydrogen ($^1H^+$; designated as hydron by the International Union of Pure and Applied Chemistry) and can disassociate/associate to Brønsted-Lowry acids/bases.[1-5] The pH is qualitatively defined as negative value of the common logarithm of the chemical activity of protons in solution and has wide-ranging implications in biological and materials sciences.[6] For example, acidification catalyzes a key part of influenza replication cycle: uncoating and intracellular release of genomic material.[7-9] Also, microorganisms have evolved genes and proteins functional in environments of extreme pH (*i.e.*, acidophiles and alkaliphiles) as opposed to normal physiological conditions near neutral pH.[10]

At a fundamental level, the pH determines the net charge on protein particles in solution relative to the isoelectric point (pI). The pI is a function of the acid dissociation equilibrium constants ($K_a$) of the R-groups of the amino acid residues comprising a protein and their proximity to solvent; $K_a$ is the ratio of dissociated acid product to associated acid reactant at thermal equilibrium. The strength of an acid is represented by the $pK_a$ ($pK_a$ = -$Log_{10}$[Ka]), where $pK_a < 0$ are strong acids (*e.g.* $pK_a^{HCl}$ = -5.9, $pK_a^{HBr}$ = -9.0, $pK_a^{HI}$ = -9.5).[11] When the $pK_a > 0$, the acid is weak and is defined as a pH buffer (*e.g.*, phosphate, tris(hydroxymethyl)aminomethane, or (4-(2-hydroxyethyl)-1-piperazineethanesulfonic acid). Characterizing the pH of biological systems at the micro- and nanoscale may have implications in assays relevant to molecular-/microbiology, virology/immunology, and other pharmaceutical and environmental sciences.[12,13]

Two-dimensional (2D) heterostructures are great candidates for measuring solutal pH because atomically thin crystals readily couple to liquid water and its ionic components at the phase interface.[14-18] For example, Wei *et al.* leveraged $Al_2O_3$/hBN nanolayers as a sensing membrane to mitigate sensor drift and for amplifying the sensitivity of monolayer $MoS_2$ field-effect transistors (FETs) as a function of solutal pH and $Al_2O_3$ thickness.[19] Furthermore, the first discovered 2D material, monolayer graphene, is comprised of elemental carbon in $sp^2$ hybridization states forming a two-dimensional electron gas (2DEG) via conjugated π-bonding networks located above and below the planar carbon crystal making it sensitive to electrostatic changes on the surface.[20,21] Additionally, the ambipolar characteristics of graphene are complemented by high-charge carrier mobility. Graphene's transfer characteristics are sensitive to dopants and charge carrier scattering events taking place on the surface through analysis of the point of minimum conduction (*i.e., the* Dirac voltage) and the transconductance, respectively.[22,23]

Graphene sensors have been demonstrated numerous times throughout the last two decades for a variety of solutal analytes and primarily measure the changes in the graphene-based devices' Dirac voltage.[24-37] For instance, recent work was done to demonstrate affinity of 3'-thiolated single stranded deoxyribonucleic acid (ssDNA) bonded to sub-percolation Au films on monolayer graphene for dilute SARS-CoV-2 RNA and other polynucleotides with a high degree of Watson-Crick complementarity to the ssDNA-S-Au/graphene/$SiO_2$.[38] Santermans *et al.* describe nonlinear screening and the pH interference mechanism of ssDNA grafting to silicon nanowires which likely has implications in future work regarding covalently bonding oligonucleotides to graphene surfaces.[39]

Graphene is ultra-sensitive to surface electrostatics and requires functionalization/defect engineering to increase sensor selectivity for analytes (*e.g.,* protons, genes, and proteins) over background noise. Measuring pH with graphene-based devices has yielded a variety of results depending on experimental conditions and device structure/fabrication method.[40-47] Typically, graphene-based devices' pH responses are highly linear and have wide-ranging sensitivity values. Recent efforts focus on functionalizing the graphene sensing mesa with acidic/basic functional groups. For example, Falina *et al.* demonstrate increased pH sensitivity of a graphene-based FET after carboxylation by anodization, whereas Bae *et al.* show that solution-deposition of $Al_2O_3$ membrane on graphene devices reliably and sensitively measures pH.[48,49]

In an alternative approach, Fu *et al.* demonstrated that atomic layer deposition of $Al_2O_3$ sensing membrane enhances pH sensitivity as compared to fluorobenzene-passivated or bare graphene alone.[50] Additionally, Fu *et al.* posit that pristine monolayer graphene is chemically insensitive to hydron and rather more sensitive to surface charge density. This may be further explicated with cation-/proton-benzene pi-complexion whereby ionized hydrogen, a quantum system with two empty electronic states, may associate and adopt electron density from the conjugated ring system comprising the graphene crystal.[51] Functionalization of 2D materials with hydroxyl-rich nanofilms, like high-κ metal oxides (*i.e.,* $Ta_2O_5$, $Al_2O_3$, HfO, or ZnO) or with chelating agents (*e.g.,* organic ionophores), enables selective detection of the chemical activity of protons in solution and necessitates further research regarding the modification of monolayer graphene surfaces.[52-58]

This work reports the pH sensing characteristics of monolayer hBN/graphene/$SiO_2$ FETs in response to 10 mM phosphate buffer titrated to varying pH values in contrast to monolayer graphene/$SiO_2$. The pH sensor scheme is outlined in **Figure 1** where **1a** is the titration of phosphate anion, a polyprotic base, through three distinct $pK_a$ values. Unique 10 mM phosphate buffered solutions of varying pH were prepared with the addition of aliquots of HCl and NaOH and were measured with an Ag/AgCl glass pH electrode. Then, the known pH solutions were incubated on the monolayer graphene/$SiO_2$ and monolayer hBN/graphene/$SiO_2$ FETs (**Figure 1a** and **1b**) and their electronic properties measured.

These FETs were derived from subtractive manufacturing of commercial, monolayer graphene/$SiO_2$ and monolayer hBN/graphene/$SiO_2$ four-inch wafers that were patterned photoresistlessly by metallizing Au electrodes with stainless steel shadow masking and electron beam deposition. Finally, the sensing mesa were defined by masking with a polypropylene microcentrifuge tube that sealed the device during reactive ion etching with argon/oxygen plasma. Polypropylene is biocompatible, making it suitable for future applications in biological assays and cell culture. This encapsulation method also sealed the source and drain Au electrodes so that the solution was unable to make contact during electrical measurements.

After repeatedly characterizing the pH sensing metrics on monolayer graphene/$SiO_2$ and monolayer hBN/graphene/$SiO_2$ FETs, the sensing mesa were coated with 9 nm of $Al_2O_3$ by electron beam deposition. The transfer characteristics' response as a function of pH was evaluated before and after developing the $Al_2O_3$ nanofilm with aqueous NaOH (pH = 12) resulting in distinct behavior. This deviation in pH-dependencies of the shift in Dirac voltage prompted a deeper study whereby Al$_2$O$_3$ was deposited on the sensing mesa by atomic layer deposition (ALD) at varying thicknesses whereby pH sensing dependencies could be evaluated accordingly. The application of 2D materials, their heterostructures, and modifications/defects for pH sensing is attractive and deserving of commercial interest.[59]

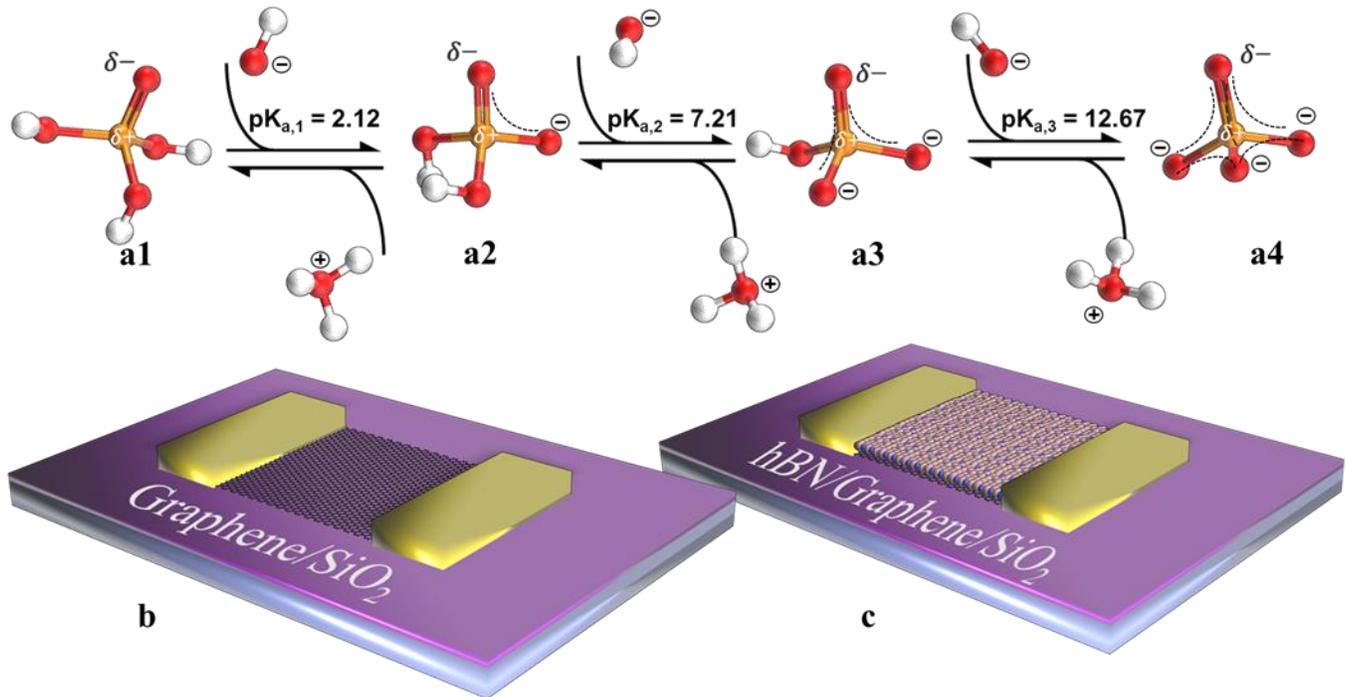

**Figure 1.** pH sensor scheme: phosphate buffered solutions of varying pH on Graphene/SiO$_2$ and hBN/Graphene/SiO$_2$. (**a1-4**) molecular model of the titration through the protonation states of phosphate anion. Three pKa values characterize three unique buffering regimes of phosphate. Dashed black lines represent resonance structures whereby electron density is delocalized. (**b**) Graphene/SiO$_2$ and (**c**) monolayer hBN/Graphene/SiO$_2$ field-effect transistors.

## Results and Discussion

### pH Sensing with Monolayer Graphene/SiO$_2$ and Monolayer hBN/Graphene/SiO$_2$

Titration of 10 mM phosphate buffer solution to pH values in the domain of [4, 10] with a step size of ΔpH = 0.5 were prepared fresh for every experiment. After reactive ion etching, the tubes were the carefully cut open forming a well where the phosphate buffered solutions could incubate on the devices' sensing mesa. Every experiment was initialized with pH = 7.0 and involved thorough washing before any electronic measurements were made: (1) add 500 μL of desired pH solution (2) remove 450 μL (3) add 450 μL of the same pH solution (4) remove 450 μL (5) add 450 μL of the same pH solution (4) remove 450 μL (5) add 150 μL of same pH solution (V$_{Final}$ = 200 μL). The devices' two- and liquid-three-terminal characteristics were then measured to extract the resistances and Dirac voltages and the washing protocol repeated as the devices were exposed to the phosphate solutions of varying pH values. Three-terminal measurements were made using a platinum electrode in the liquid.

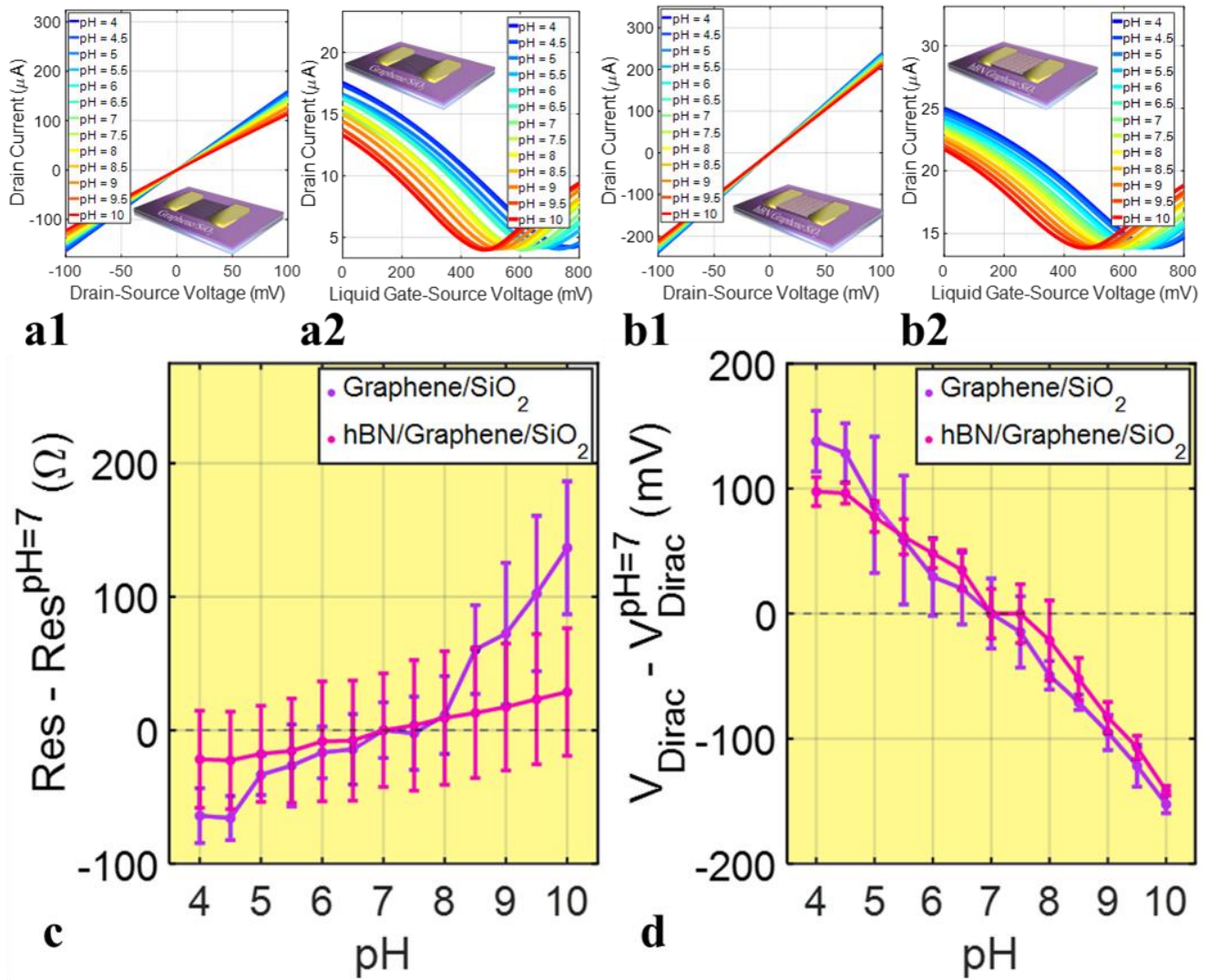

**Figure 2:** Two- and three-terminal characteristics on single (**a1-2**) graphene/SiO$_2$ and (**b1-2**) monolayer hBN/graphene/SiO$_2$ FETs in response to 200 μL of 10 mM phosphate buffers of varying pH. Three-terminal characteristics were liquid-gated with a platinum electrode. The average change in resistance (**c**) and Dirac voltage (**d**) relative to the resistance and Dirac voltage measured at pH = 7. The standard deviation was calculated from n = 3 replicates on each device.

This washing protocol was employed strictly to avoid drying the devices' surfaces completely to mitigate any deposition of salt residue on the 2D materials' sensing mesa. Additionally, this washing method ensured that the pH on the device was the pH originally measured with the Ag/AgCl glass electrode when the phosphate buffer solutions were originally prepared. The washing protocol was verified by remeasuring the pH values of the final volume (200 μL) 10 mM phosphate solutions used for the electronic measurements. The measurements were initialized with pH = 7 and first stepped into the acidic regime. After stepping to, and measuring, pH = 4, the devices was reinitialized with pH = 7 and then stepped into the basic regime (up to pH = 10).

This was repeated in triplicate on each device to yield the results displayed in **Figure 2**. The raw two- and liquid-three-terminal data for monolayer graphene/SiO$_2$ is displayed in **2a1** and **2a2**, whereas monolayer hBN/graphene/SiO$_2$ is displayed in **2b1** and **2b2.** The changes in resistances and Dirac voltages relative to the measured values at pH = 7 (reinitialized) are shown in **Figure 2c** and **2d**.

In **2c**, the resistances of monolayer hBN/graphene/SiO$_2$ has higher linearity, but much larger standard deviation, as a function of pH. The standard deviations of the monolayer graphene/SiO$_2$ FETs are lower in the acidic regime as compared to the basic and have a larger range than the hBN/graphene/SiO$_2$. Both devices demonstrate positive correlation between pH and resistance. This is further corroborated in **Figure 2d**, which shows a negative correlation between Dirac voltage and pH, which contrasts to much of the previous literature and has recently been hypothesized by Lee *et al* to be due to the electrostatic effects that are a function of phosphate molarity.[60] Here, both graphene- and hBN/graphene-based devices become more p-type as the pH becomes acidic. At high pH, phosphate is more anionic (*i.e.,* more net-negatively charged) as is depicted in **Figure 1a4** and may couple to the p-type carriers in the graphene channels resulting in the devices becoming less p-type.

Overall, the monolayer hBN/graphene/SiO$_2$ device had both a lower range and standard deviation as compared to the monolayer graphene/SiO$_2$ device. Interestingly, the changes in Dirac voltages for the graphene/SiO2 device were lower in the basic regime, which starkly contrasts to the resistances' trend as a function of the pH. The Dirac voltage depends on the doping level of the graphene-based devices and is more sensitive to changes in surface charge density than the resistance, which is a function of not only the doping level, but also charge carrier mobility. In the next sections of this report, the liquid-transfer characteristics of (1) electron beam deposition and (2) atomic layer deposition of Al$_2$O$_3$ nanofilms on monolayer graphene/SiO$_2$ and monolayer hBN/graphene/SiO$_2$ are explored to contrast the deposition methods and characterize the dependence of the pH sensing properties on the thickness of Al$_2$O$_3$.

### *pH Sensing with Electron Beam Deposition of 9 nm of Al$_2$O$_3$ on Monolayer Graphene/SiO$_2$ and Monolayer hBN/Graphene/SiO$_2$*

Functionalization of the monolayer graphene/SiO$_2$ and monolayer hBN/graphene/SiO$_2$ with 9 nm of Al$_2$O$_3$ by electron beam deposition (thickness gauged from quartz crystal resonator) showed significant aberration to the bare devices displayed in **Figure 2**. The devices were characterized with Raman spectral mapping (**Figure S2**) and are shown to have comparable 2D/G phonon intensity ratios; however, the 9 nm Al$_2$O$_3$/graphene/SiO$_2$ D/G ratios were, on average, larger than that of the 9 nm Al$_2$O$_3$/hBN/graphene/SiO$_2$. This increase in D/G may signify that the electron beam deposition of Al₂O₃ alters the quality of the graphene material (reference Raman mapping spectra for fresh, bare monolayer graphene/SiO2 can be found in **Figure S3a1** for comparison)

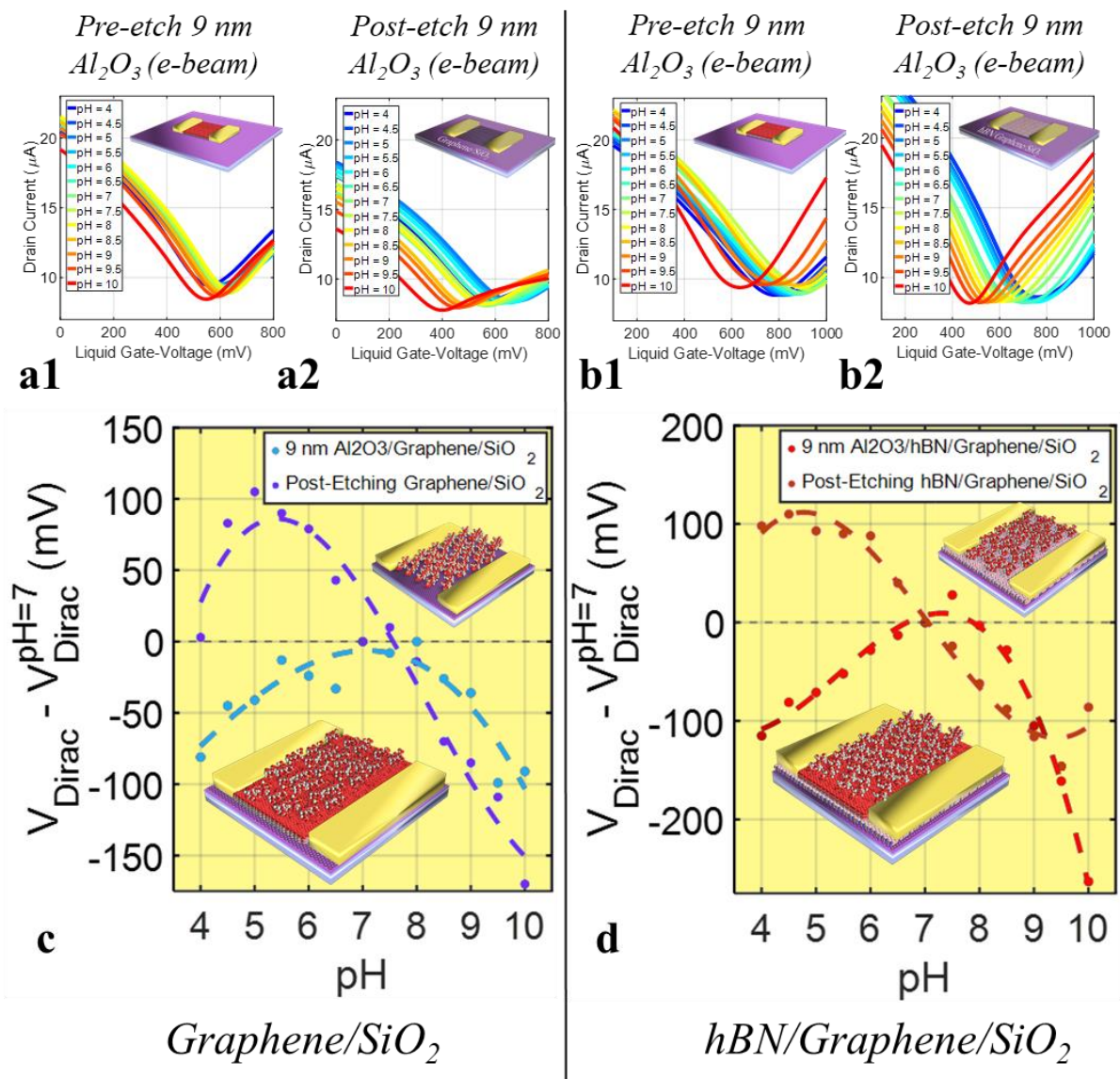

**Figure 3**. Liquid-three-terminal characteristics as a function of 10 mM phosphate pH before and after etching 9 nm of Al₂O₃ (electron beam deposition) from **(a1-2)** graphene/SiO₂ and **(b1-2)** monolayer hBN/graphene/SiO₂ (insets show device structure). Shifts in the Dirac voltage relative to the Dirac voltage at pH = 7 are plotted for graphene/SiO₂ **(c)** and monolayer hBN/graphene/SiO₂ **(d)** at the various pH values before and after etching the Al₂O₃. **(c, d)** insets depict phosphate buffered solution on the various device sensing mesa.

The pH sensing results from the electron beam deposition are shown in **Figure 3**, where the devices were exposed to 10 mM phosphate buffered solutions with the same protocol as described above. After exposing the devices to the pH solutions, the Al₂O₃-functionalized devices were incubated in pH = 12 NaOH(aq) to develop the Al₂O₃ from the surface by wet etching. Leveraging the solubility of Al₂O₃ in highly-basic solution provides a way to characterize the effects of the Al₂O₃ nanofilm on the pH sensing-properties of the graphene/$SiO_2$ and hBN/graphene/$SiO_2$ FETs by studying the deviations from pristine, bare surfaces. In past literature, 1-5% tetramethylammonium hydroxide (TMAH)-based developers (pH ≈ 13) has been used for etching $Al_2O_3$ from graphene.[62] The liquid-transfer characteristics' raw data is shown for graphene/$SiO_2$ and hBN/graphene/$SiO_2$ before (**2a1** and **2b1**) and after (**2a2** and **2b2**) basic etching, respectively.

The change in Dirac voltages as a function of pH, relative to the Dirac voltage measured at pH = 7 (reinitialized) is summarized before and after the basic etch of 9 nm $Al_2O_3$/monolayer graphene/$SiO_2$ (**2c**) and 9 nm $Al_2O_3$/monolayer hBN/graphene/$SiO_2$ (**2d**). Before the basic etch, both graphene/$SiO_2$ and hBN/graphene/$SiO_2$ devices showed a concave-down response centered near pH = 7 and pH = 7.5, respectively, where the hBN/graphene/$SiO_2$ device response had greater curvature. With the electron beam deposition of 9 nm of $Al_2O_3$, both devices showed decreases in the Dirac voltage about the centered-pH. Removal of the $Al_2O_3$ with wet etching adjusted the pH responses of the graphene/$SiO_2$ and hBN/graphene/$SiO_2$ devices to resemble the responses measured in **Figure 2**, however, the response was not completely reverted.

Specifically, after basic wet etching, the graphene/$SiO_2$ device showed negative correlation with linearity within the domain of pH = [5, 10], whereas the hBN/graphene/$SiO_2$ device showed negative correlation within a smaller domain of pH = [6, 9.5]. The total ranges of change in Dirac voltage, post-etching of the $Al_2O_3$, of the graphene/$SiO_2$ and hBN/graphene/$SiO_2$ devices were like the ranges measured in **Figure 2**.

The aberration in pH-response with the $Al_2O_3$ nanofilm may be due to the deposition method, whereby an electron beam is used to sublime $Al_2O_3$ by increasing the temperature under vacuum. This method is high-energy and may result in morphological changes in the $Al_2O_3$ when deposited on the surface of graphene/$SiO_2$ and hBN/graphene/$SiO_2$. Furthermore, Sun *et al.* shows the solubility of $Al_2O_3$ on ZnO increases with increasing the pH of solution and may begin to etch the nanofilm at pH > 8.[61] More work is needed to characterize the etching rate of $Al_2O_3$ versus pH of $NaOH_{(aq)}$ on monolayer graphene/$SiO_2$ or hBN/graphene/$SiO_2$ and how the etching alters the surface and, ultimately, the charge transport characteristics.

The limited reversion of the pH response after basic wet etching may result from residual, unetched $Al_2O_3$ remaining on the sensing mesa surface. Additionally, the high-energy deposition method may increase the crystallinity of the $Al_2O_3$ nanofilm, resulting in an altered etching rate in basic conditions. These results prompted another study using atomic layer deposition, a well-controlled surface deposition method that relies on amorphous solidification of $Al_2O_3$ on the graphene/$SiO_2$ and hBN/graphene/$SiO_2$ devices.

### pH Sensing with Atomic Layer Deposition of Varying Thicknesses of $Al_2O_3$ on Monolayer Graphene/$SiO_2$ and Monolayer hBN/Graphene/$SiO_2$

Atomic layer deposition (ALD) was employed to contrast the dependency of the pH sensing properties of graphene- and hBN/graphene-based devices on the deposition method of $Al_2O_3$, the thickness of $Al_2O_3$, and the capacity for the sensors to revert their response akin to the fresh, bare graphene/$SiO_2$ and hBN/graphene/$SiO_2$ devices characterized in **Figure 2**. The ALD of $Al_2O_3$ of varying thicknesses were measured with a crystal quartz resonator during deposition, and ellipsometry after deposition. The measurements of the thicknesses of $Al_2O_3$ in this study are shown in **Table I**.

The, the $Al_2O_3$ on both graphene/$SiO_2$ and hBN/graphene/$SiO_2$ were subjected to Raman spectral mapping (n = 625 spectra) and atomic force microscopy (AFM). From the data, average 2D/G and D/G phonon intensity ratios were extracted and plotted against the thickness of $Al_2O_3$ on the surface (**Figures 5a** and **5b**). The Raman spectral maps can be found in **Figure S3**. In general, the monolayer graphene/$SiO_2$ had a higher 2D/G ratio up to $t_4$, correspondent to larger in-plane vibrations of the graphene crystal. The hBN/graphene/$SiO_2$ 2D/G ratios, while lower than the graphene/$SiO_2$ devices, had a smaller standard deviation and is correspondent to greater planar uniformity of the graphene material.

**Table I. $Al_2O_3$ Thickness Metrology**

| Thickness Index | Quartz Crystal Resonator Thickness of $Al_2O_3$ (nm) | Ellipsometer Thickness of $Al_2O_3$ (nm) |
| --- | --- | --- |
| $t_1$ | - | - |
| $t_2$ | 3.6 | 4.4 |
| $t_3$ | 7.2 | 8.6 |
| $t_4$ | 10.8 | 12.3 |
| $t_5$ | 23.0 | 23.0 |

**Table I**. Atomic layer deposition of $Al_2O_3$ thicknesses as measured by the quartz crystal resonator and ellipsometry.

Interestingly, 23 nm of $Al_2O_3$ shows aberration in the trend of 2D/G and D/G phonon intensity ratios and may be correlated to coalescence of the $Al_2O_3$ nanofilm. In **5b**, the hBN/graphene/$SiO_2$ had, on average, a lower D/G phonon intensity ratio, suggesting a lower degree of defects being present, however, the standard deviations large. The D/G ratios are much lower for the ALD method as compared to the electron beam deposition. This may be related to strain on the 2D crystals induced by the nanofilm. Atomic force microscopy (AFM) was then employed to characterize the surfaces.

The root-mean-square (RMS) roughness as a function of the thickness of $Al_2O_3$ is shown in **Figure 5c**. The roughness of bare hBN/graphene/$SiO_2$ is larger than bare graphene/$SiO_2$ and the raw AFM data can be found in **Figure S4**. Interestingly, deposition of $Al_2O_3$ on hBN/graphene/$SiO_2$ results in a decrease in RMS roughness up to 8.6 nm of $Al_2O_3$, where greater thicknesses show the RMS roughness remains relatively consistent. This is hypothesized to be due to coalescence of the $Al_2O_3$ and can be observed in **Figure S4g-j**. Interestingly, the graphene/$SiO2$ samples display a notable coalescence as the $Al_2O_3$ thickness increases and a slight increase in the RMS roughness.

After characterizing the thicknesses, Raman spectral qualities, and the AFM topography, the devices were processed with microcentrifuge tube encapsulation and subjected to exposure to 10 mM phosphate buffered solution of varying pH. In this study, to conserve time and expedite experimentation, the pH interval used was $\Delta pH = 1.0$ within the domain of pH = [4, 10]. The primary findings are outlined in **Figure 6**. The freshly deposited $Al_2O_3$/graphene/$SiO_2$ and $Al_2O_3$/hBN/graphene/$SiO_2$ devices had their pH responses characterized (**Figures 6a** and **6b**, respectively) which elucidated dependencies on the pH sensing properties on the thickness of ALD Al₂O₃ that were again distinct to the response of the bare 2D materials in **Figure 2d.**

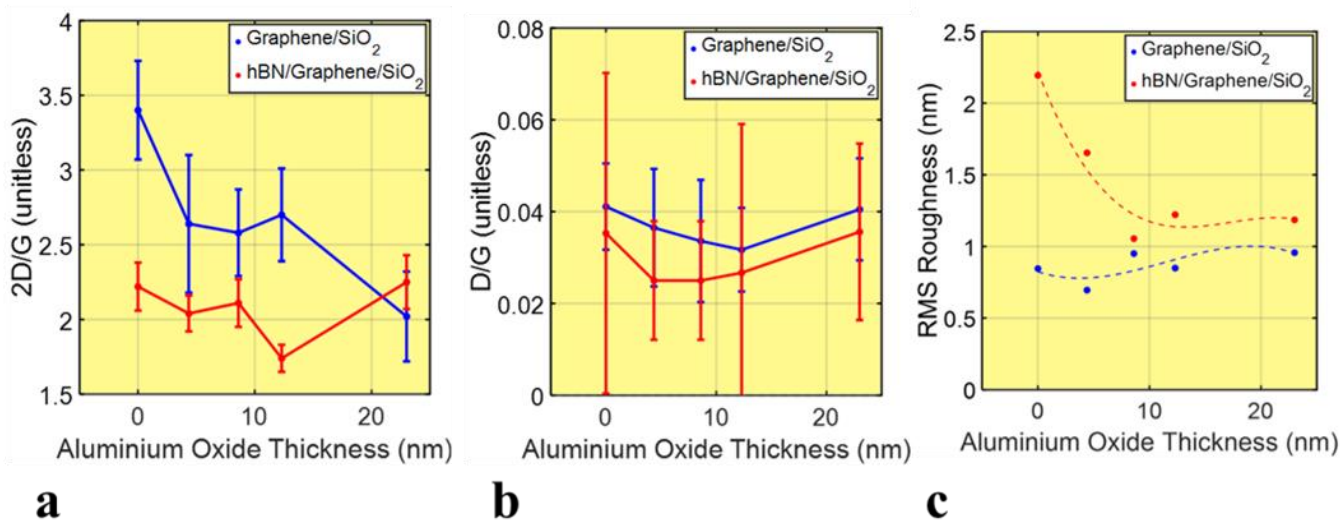

**a**                                       **b**                                       **c**

**Figure 5**. (**a**) Monolayer graphene/SiO₂ and monolayer hBN/graphene/SiO₂ 2D/G phonon intensity average ratios and their standard deviations (n = 625 spectra) and (**b**) their D/G phonon intensity average ratios and their standard deviations. (**c**) Atomic force microscopy of the RMS roughness versus the thickness of Al₂O₃ on monolayer graphene/SiO₂ and monolayer hBN/graphene/SiO₂.

In **Figure 6a,** the reference sample (*i.e.,* 0 nm Al₂O₃) has a slightly smaller range of shift in Dirac voltage and similar linearity as compared to **Figure 2d**. Depositing 4.3 nm of Al₂O₃ on graphene/SiO₂ exhibited a lower domain of linearity from pH = [5, 10] as compared to the reference which showed high linearity from pH = [4, 10]. Deposition of 8.6 nm of Al₂O₃ further altered the domain of linearity from pH = [7, 10] with negligible sensitivity in the acidic regime and an enhanced sensitivity in the basic regime. The Al₂O₃ sample (12.3 nm) had its domain of linearity further reduced to pH = [8, 10]. Additionally, for 12.3 nm Al₂O₃/graphene/SiO₂, the sensitivity in the acidic regime is small and positively correlated, which contrasts sharply with the response to acidic 10 mM phosphate buffer of the reference sample.

Interestingly, the domain of linearity for Al₂O₃/hBN/graphene/SiO₂ in **6b** remains relatively constant for pH = [5, 10] for all thickness of Al₂O₃ as compared to the reference sample (*i.e.*, 0 nm Al₂O₃/hBN/graphene/SiO₂). The reference sample showed similar pH sensing response as the data shown in **2d**. Furthermore, the sensitivity and range of the Dirac voltage in response to the pH of 10 mM phosphate buffer solutions is maximized for the 8.6 nm Al₂O₃/hBN/graphene/SiO₂ devices and shows a sharp reduction when the thickness is increased to 12.3 nm of Al₂O₃. The 4.3 nm Al₂O₃/hBN/graphene/SiO₂ sample had intermediate sensitivity and range. The liquid-transfer characteristics raw data can be found in **Figure S5.**

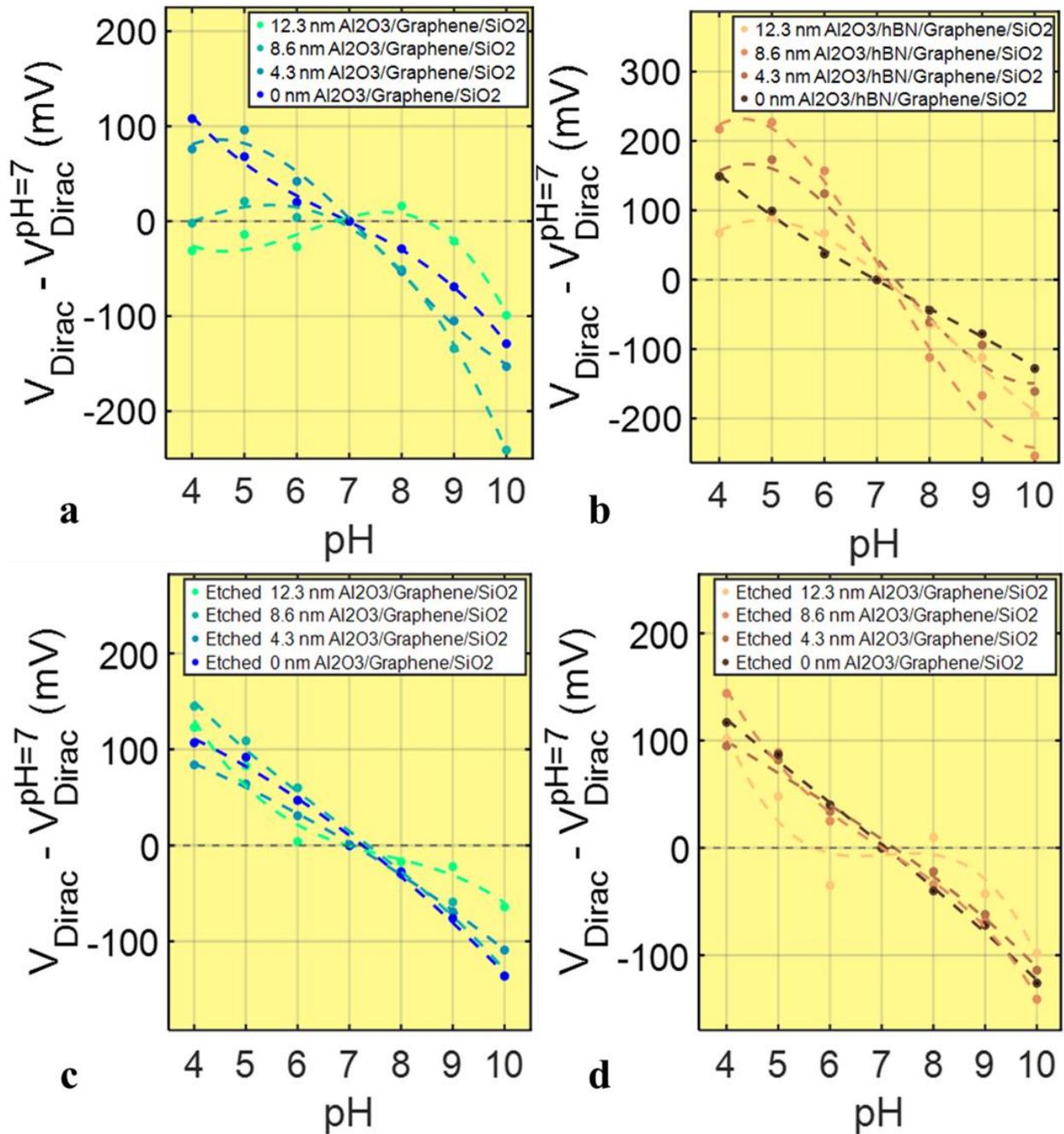

**Figure 6**. Dirac voltages of varying thickness of ALD $Al_2O_3$ on graphene/$SiO_2$ (**a**) and monolayer hBN/graphene/$SiO_2$ (**b**) as a function of 10 mM phosphate pH and after basic etching (**c, d**), respectively. Thickness of $Al_2O_3$ used here are: 0 nm, 4.3 nm, 8.6 nm, and 12.3 nm.

To further integrate the dependencies of the deposition method of $Al_2O_3$, the devices were subjected to basic etch with $NaOH_{(aq)}$ (pH = 12) immediately after characterizing the pH response and the experiment was repeated. The post-etching results, shown in **6c** and **6d**, suggest that the pH sensing properties of the devices can be reverted to the bare 2D materials pH sensing properties up to 8.6 nm of ALD nm $Al_2O_3$. Intriguingly, the 12.3 nm $Al_2O_3$ on both graphene/$SiO_2$ and on

hBN/graphene/SiO$_2$ show aberration from the reference FETs, with the hBN/graphene/SiO$_2$ device having a more distinct change. Specifically, two domains of negative correlation are found between pH = [4, 6] and [8, 10] and a domain of positive correlation from pH = [6, 8] for the hBN/graphene/SiO$_2$ device originally coated with 12.3 nm of Al$_2$O$_3$.

### Table II. Device Sensitivity Summary

| Sensing Mesa | Al$_2$O$_3$ Thickness | Domain(s)-of-Linearity (pH) | Linear Sensitivity (mV/pH) | R$^2$ |
|---|---|---|---|---|
| Graphene/SiO$_2$ | 0 nm | [4, 10] | -39.2 | 0.9553 |
| | | [4, 10] | -49.0 | 0.9887 |
| | | [4, 10] | -53.6 | 0.9873 |
| | Pre-etch 9 nm (e-beam) | [4, 8] | 16.6 | 0.7491 |
| | | [8, 10] | -51.0 | 0.8843 |
| | Post-etch 9 nm (e-beam) | [4, 5] | 102 | 0.9027 |
| | | [5 ,10] | -52.8 | 0.9687 |
| | Pre-etch (-)-Al$_2$O$_3$ Control (0 nm) | [4, 10] | -36.9 | 0.9843 |
| | Post-etch (-)-Al$_2$O$_3$ Control (0 nm) | [4, 10] | -40.7 | 0.9839 |
| | Pre-etch 4.4 nm (ALD) | [4, 5] | 20 | 1 |
| | | [5,10] | -49.7 | 0.9992 |
| | Post-etch 4.4 nm (ALD) | [4, 10] | -32.3 | 0.9910 |
| | Pre-etch 8.6 nm (ALD) | [5, 7] | -2.6 | 0.5744 |
| | | [7, 10] | -80.6 | 0.9763 |
| | Post-etch 8.6 nm (ALD) | [4, 10] | -45.3 | 0.9873 |
| | Pre-etch 12.3 nm (ALD) | [4, 8] | 10.8 | 0.77 |
| | | [8, 10] | -57.5 | 0.9594 |
| | Post-etch 12.3 nm (ALD) | [4, 10] | -28.3 | 0.8866 |
| hBN/Graphene/SiO$_2$ | 0 nm | [4, 10] | -36.7 | 0.8960 |
| | | [4, 10] | -40.1 | 0.9863 |
| | | [4, 10] | -42.2 | 0.9846 |
| | Pre-etch 9 nm (e-beam) | [4, 7.5] | 38.2 | 0.9887 |
| | | [7.5, 10] | -114.6 | 0.9499 |
| | Post-etch 9 nm (e-beam) | [4, 6] | -8 | 0.5181 |
| | | [6, 9.5] | -64.8 | 0.9914 |
| | Pre-etch (-)-Al$_2$O$_3$ Control (0 nm) | [4, 10] | -45.2 | 0.9938 |
| | Post-etch (-)-Al$_2$O$_3$ Control (0 nm) | [4, 10] | -40.3 | 0.9967 |
| | Pre-etch 4.4 nm (ALD) | [5, 10] | -68.1 | 0.9706 |
| | Post-etch 4.4 nm (ALD) | [4, 10] | -34.7 | 0.9844 |
| | Pre-etch 8.6 nm (ALD) | [5, 10] | -99.7 | 0.9801 |
| | Post-etch 8.6 nm (ALD) | [4, 10] | -43.9 | 0.9808 |
| | Pre-etch 12.3 nm (ALD) | [5, 10] | -57.9 | 0.9835 |
| | Post-etch 12.3 nm (ALD) | [4, 6] | -68 | 0.9840 |
| | | [6, 8] | 22.5 | 0.9067 |
| | | [8, 10] | -54 | 0.9999 |

**Table II**. Sensitivities of the graphene/SiO$_2$ and hBN/graphene/SiO$_2$ devices with/out Al$_2$O$_3$ nanofilm.

The deviation observed for the 12.3 nm $Al_2O_3$ samples may be due to residual $Al_2O_3$ that was incompletely etched but requires destroying the device to investigate the surface further. The etching of $Al_2O_3$ from the devices was monitored with chronoamperometry (**Figure S6**) and shows correlation in time to stabilization as a function of increased thickness of $Al_2O_3$ and, generally, a different response than the bare 2D materials. A summary of the domains-of-linearity and the linear sensitivities of the devices used in this work is shown in **Table II**.

**Conclusions**

In this work, graphene- and hBN/graphene-based FETs were used for measuring pH. Initially, pH responses of bare graphene/$SiO_2$ and hBN/graphene/$SiO_2$ FETs were characterized with 10 mM phosphate buffer, and the liquid-transfer characteristics proved to be a more reliable measurand than the device resistance. In this scenario, the graphene/$SiO_2$ and hBN/graphene/$SiO_2$ FETs had a negative correlation to increasing pH (**Table II**). Both devices became less p-type for basic pH values and may be related to the electron-rich structure of phosphate anion coupling with holes in the graphene valence band and may have been partially screened by the hBN capping layer resulting in the slightly reduced range observed in **Figure 2d**. The bare hBN/graphene/$SiO_2$ devices were more reliable than the graphene/SiO2 devices with a lower standard deviation. The response of the hBN/graphene/$SiO_2$ in the acidic and basic regimes were distinct and is hypothesized to be due to incommensurate screening of solutal ions by the insulating monolayer hBN, meriting further research.

Electron beam deposition of 9 nm $Al_2O_3$ on graphene/$SiO_2$ and hBN/graphene/$SiO_2$ altered the pH response markedly, demonstrating a concave-down correlation centered near neutral pH values. Electron beam deposition of $Al_2O_3$ is expected to yield a more crystalline nanofilm than ALD. The higher crystallinity may have larger net polarization, forming a strong dielectric layer on the 2D material surface that influences pH sensing properties.

The ALD $Al_2O_3$ showed reversion to the bare 2D materials response up to 12.3 nm upon basic wet etching. Incomplete reversion of the pH sensing properties is not fully understood, but may be due to residual, unetched $Al_2O_3$ and deserves future work. Additionally, an $Al_2O_3$ thickness-dependent pH response was observed, showing the 8.6 nm of $Al_2O_3$/hBN/graphene/$SiO_2$ was the most sensitive device studied here with super-Nernstian sensitivity of -99.7 mV/pH within the domain of pH = [5, 10], a phenomena that has recently been demonstrated by Jung *et al.* to be based on anomalous charge transfer and defect engineering of graphene crystal grain boundries.[63]

Functionalization of graphene and hBN/graphene FETs with $Al_2O_3$ has been shown to alter the pH response to 10 mM phosphate buffered solutions depending on the deposition method. The reversion of the pH response after basic wet etching demonstrates the function of the $Al_2O_3$ nanofilm. Future work should detail the effects of different metal oxide nanofilms (*e.g.*, $Ta_2O_5$, HfO, or ZnO) and explore the incorporation of proton ionophore membranes. The work outlined here suggests that hBN/graphene heterostructures can be used as reliable pH sensors with pH responses that are more tunable as a function of thickness of $Al_2O_3$ nanofilms than on monolayer graphene/$SiO_2$ alone.

**Methods and Materials**

*Reagents*

Four-inch monolayer graphene/90 nm of $SiO_2$/p-type silicon and monolayer hBN/graphene/90 nm of $SiO_2$/p-type silicon was commercially sourced from Grolltex. Shadow masks were stainless-steel and acquired from OSH Stencils. Reagents were bought from well-known vendors. All water used was molecular biology-grade deionized water and filtered. Monobasic and dibasic sodium phosphate and aqueous 1.0 M HCl and NaOH were obtained from Millipore Sigma. Au and $Al_2O_3$ were purchased from Kurt J. Lesker for the electron beam depositions. Thermo Scientific Orion™ 9810BN Micro pH Ag/AgCl glass microelectrode was used to measure the pH of the 10 mM phosphate buffered solutions.

*Graphene/$SiO_2$ and hBN/Graphene/$SiO_2$ Field-Effect Transistor Fabrication*

Monolayer graphene/ four-inch 90 nm of $SiO_2$/p-type silicon and monolayer hBN/graphene/90 nm of $SiO_2$/p-type silicon was used. Stainless-steel shadow masking was used for the metallization with 100 nm of Au via electron beam deposition with Angstrom deposition system. The wafer was then cleaved with a diamond tip pen for chip-scale processing. Microcentrifuge tube encapsulation defined the sensing area of $\approx 10.5$ mm$^2$ (radius = 3.25 mm ; area = $\pi r^2$) by trimming the cap of the microcentrifuge tube and using a UV-curable adhesive to bond it to the device surface. Then, the encapsulated device was subjected to reactive ion etching with argon/oxygen plasma. After dry etching, the microcentrifuge tube tip was trimmed, opening the volume used to analyze the various phosphate buffered solutions.

*Electrical Measurements*

Electrical measurements were acquired by connecting the device through a probe station to a Keysight B2902A precision source meter. Every three-terminal measurement was done with a drain-source bias of 10 mV. Liquid-gated transfer characteristics were in the domain and polarity of [-400, +1000] mV and sometimes extended to [-400, +1200] mV when the Dirac Voltage was greater than 1000 mV. The measurements used 20 power line cycles. Liquid-gated measurements were made with a clean platinum electrode.

*Raman Spectroscopy*

Horiba Raman microscope was employed for generating Raman spectral maps from 625 separate spectra over 62.5 x 62.5 μm$^2$ ($L \times W$) area. Spatial discretization between measurements was 2.5 μm in both principal directions describing the surface plane. Measurement parameters are as follows: excitation wavelength of 532 nm, exposure time of 5 seconds and 4 accumulations per point with a power of 14 mW. A grating of 1200 I/mm was used, and the spectrum centered at 1800 cm$^{-1}$.

**Acknowledgments**

The authors are appreciative and recognize the funding provided by Analog Devices, Inc. (Award no. 9300011743) and Boston University's optoelectronic processing, and characterization facilities. Furthermore, Grolltex, the supplied of monolayer graphene and monolayer hBN/graphene on four-inch SiO2/Si wafers, is deserving of special mention for making this study possible.

## Supporting Information Available

Raw data for liquid-transfer characteristics, Raman spectral mapping of electron beam and atomic layer deposition of $Al_2O_3$ of varying thicknesses, atomic force microscopy topography of the $Al_2O_3$ nanofilms, and chronoamperometry of the basic etching of $Al_2O_3$ from the surface of the 2D materials used as the active mesa for pH sensing.